%% file: prl_CMLLP_V81.tex
\documentclass[english,aps,prl,twocolumn,showpacs,superscriptaddress,groupedaddress,amsmath,amssymb,nofootinbib]{revtex4}
\usepackage[T1]{fontenc}
\usepackage[latin9]{inputenc}
\usepackage{verbatim}
\usepackage{float}
\usepackage{amsmath}
\usepackage{graphicx}
\usepackage{amssymb}

\makeatletter
\@ifundefined{textcolor}{}
{%
 \definecolor{BLACK}{gray}{0}
 \definecolor{WHITE}{gray}{1}
 \definecolor{RED}{rgb}{1,0,0}
 \definecolor{GREEN}{rgb}{0,1,0}
 \definecolor{BLUE}{rgb}{0,0,1}
 \definecolor{CYAN}{cmyk}{1,0,0,0}
 \definecolor{MAGENTA}{cmyk}{0,1,0,0}
 \definecolor{YELLOW}{cmyk}{0,0,1,0}
 }

\usepackage{graphics}
\usepackage{mathptmx}
\usepackage{dcolumn}
\usepackage{bm}
\makeatother

\usepackage{babel}

\begin{document}


\hspace{5.2in} \mbox{FERMILAB-PUB-11-534-E}

\renewcommand{\thefootnote}{\fnsymbol{footnote}} 

\title{A search for charged massive long-lived particles}

\input author_list.tex       
\date{October 12, 2011}

\begin{abstract}
We\textsc{ }report on a search for charged massive long-lived particles
(CMLLPs), based on 5.2 fb$^{-1}$ of integrated luminosity collected
with the D0 detector at the Fermilab Tevatron $p\bar{p}$ collider.
We search for events in which one or more particles are reconstructed
as muons but have speed and ionization energy loss $(dE/dx)$ inconsistent
with muons produced in beam collisions. CMLLPs are predicted in several
theories of physics beyond the standard model. We exclude pair-produced
long-lived gaugino-like charginos below 267 GeV and higgsino-like
charginos below 217 GeV at 95\% C.L., as well as long-lived scalar
top quarks with mass below 285 GeV.
\end{abstract}

\maketitle

We report on a search for massive particles, which are electrically
charged and have a lifetime long enough to escape the D0 detector
before decaying. Charged massive long-lived particles are
not present in the standard model (SM) nor are their distinguishing
characteristics (slow speed, high $dE/dx$) relevant for most high energy physics
studies. Although the distinctive signature in itself provides sufficient motivation for
a search, some recent extensions to the SM suggest that CMLLPs exist
and are not yet excluded by cosmological limits$\,$\cite{cosmo1,cosmo2}.
Indeed, the standard model of big bang nucleosynthesis (BBN) has difficulties in explaining the observed lithium production.
The existence of a CMLLP that decays during or after the time of BBN
could resolve this disagreement \cite{BBN}. %
{}

We derive cross section limits for CMLLPs and compare them to theories
of physics beyond the SM.  In most supersymmetric (SUSY) models the lightest SUSY particle is assumed to be stable.  Some SUSY models predict that the next-to-lightest supersymmetric particle (NLSP) can be a CMLLP.  In this Letter we explore models which include a  chargino as a NLSP whose lifetime
can be long if its mass differs from the mass of the lightest neutralino
by less than about 150 MeV$\,$\cite{AMSB1,AMSB2}. This can occur
in models with anomaly-mediated supersymmetry breaking (AMSB) or in
models that do not have gaugino mass unification. There are two general
cases, where the chargino is mostly a higgsino and where the chargino
is mostly a gaugino, which we treat separately in this Letter. 

There are some SUSY models that predict a scalar top quark (stop)
NLSP and a gravitino LSP. These stop quarks hadronize into both charged
and neutral mesons and baryons which live long enough to be CMLLP
candidates \cite{GMSB3}. Further, hidden valley models predict GMSB-like
scenarios where the stop quark acts like the LSP and does not decay
but hadronizes into charged and neutral hadrons that escape the detector $\,$\cite{HiddenValley1,HiddenValley2}.
In general, any SUSY scenario where the stop quark is the lightest
colored particle (which will happen in models without mass unification
and heavy gluinos) can have a stop CMLLP. Any colored CMLLP will undergo hadronization and charge exchange during
nuclear interactions, which we discuss below.

This search utilizes data collected between 2006 and 2010 with the
D0 detector \cite{d0det} at Fermilab's 1.96 TeV $p\bar{p}$ Tevatron
Collider, and is based on 5.2 fb$^{-1}$ of integrated luminosity.
We reported earlier \cite{D0 MSP PRL} on a similar 1.1 fb$^{-1}$
study, searching for events with a pair of CMLLPs, each with low speed.  In addition to using the larger data sample, the present
search looks for one or more CMLLP, rather than only for a pair, and characterizes CMLLPs  with high $dE/dx$ in addition to slow speed.  Other searches for long-lived particles include
those from the CDF Collaboration $ $$\,$\cite{CDF Run1 PRL,CDF Run2 PRL},
the CERN $e^{+}e^{-}$$\,$Collider LEP $\:$\cite{LEP searches},
and the CERN $pp\,$ Collider LHC \cite{CMS HSCP paper,ATLAS cmllp paper}.

%
{}

The D0 detector$\,$\cite{d0det} includes an inner tracker with two
components: an innermost silicon microstrip tracker (SMT) and a scintillating
fiber detector. We find a particle's $dE/dx$ from the energy losses
associated with its track in the SMT. The tracker is embedded within
a 1.9 T superconducting solenoidal magnet. Outside the solenoid is a
uranium/liquid-argon calorimeter surrounded by a muon spectrometer,
consisting of drift tube planes on either side of a 1.8 T iron toroid.
There are three layers of the muon detector: the A-layer, located
between the calorimeter and the toroid, and the B- and C- layers,
outside the toroid. Each layer includes scintillation counters
which serve to veto cosmic rays. Thus the muon system provides multiple
time measurements from which a particle's speed may be calculated.

\begin{figure}[]
\noindent \begin{centering}
\includegraphics[scale=0.47]{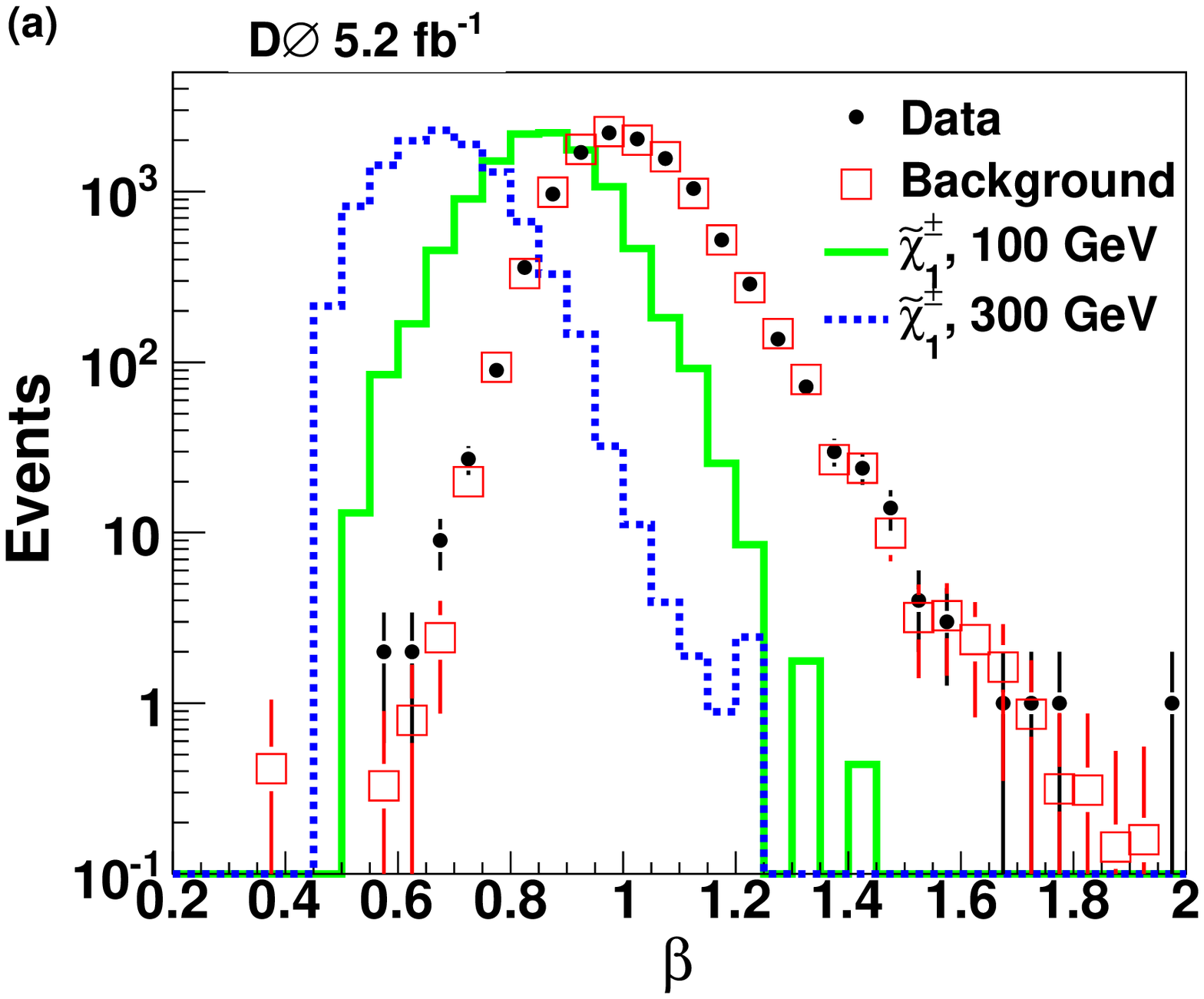}
\par\end{centering}

\noindent \begin{centering}
\includegraphics[scale=0.47]{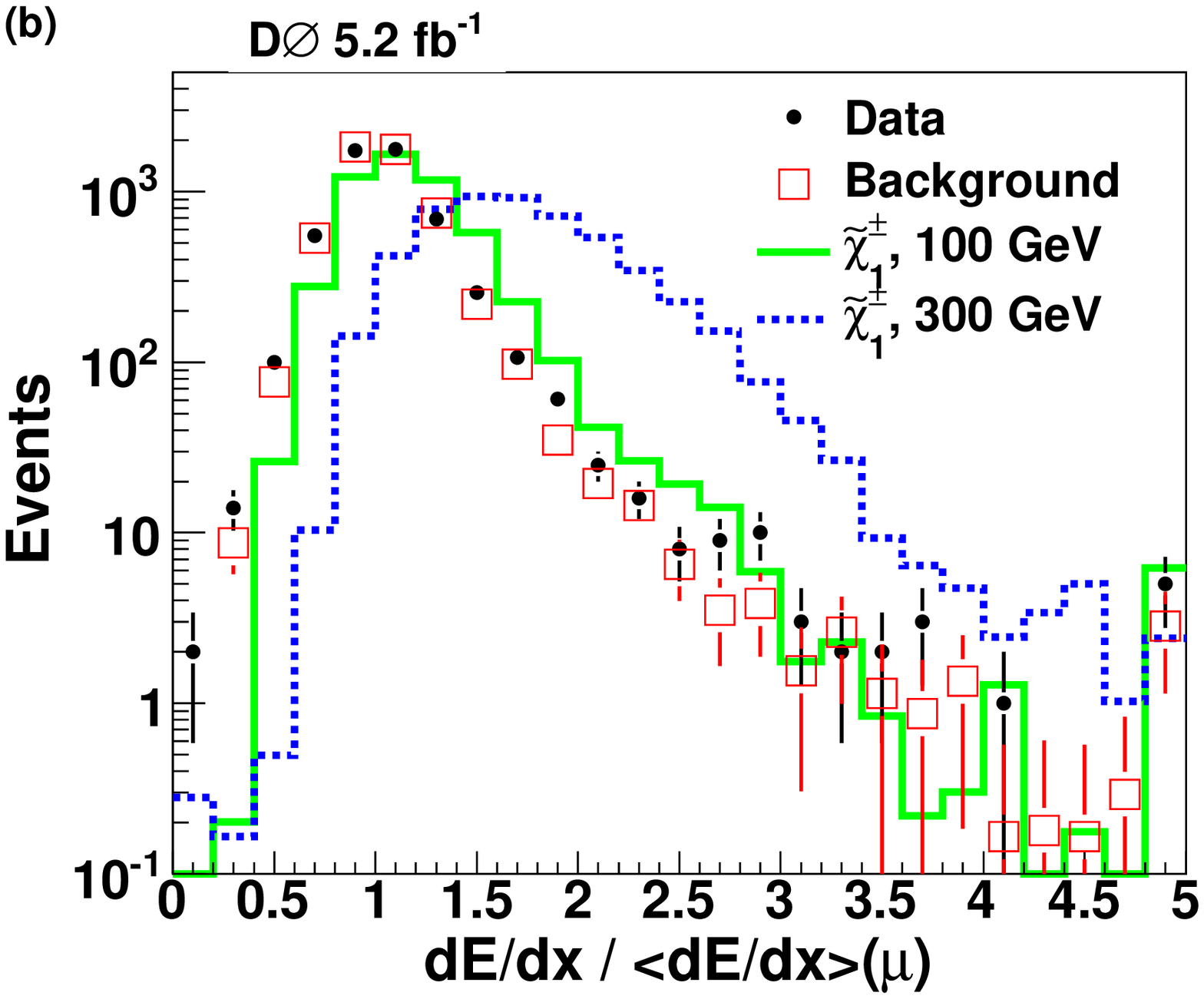}
\par\end{centering}

\caption{\label{fig:beta_dedx}Distributions of (a) speed$\:\beta$ and (b)
$dE/dx$ for data, background, and signal (gaugino-like charginos
with a mass of 100 and 300 GeV) that pass the selection criteria.
The histograms have been normalized to have the same numbers of events.
We have adjusted the scale of the $dE/dx$ measurements so that the
$dE/dx$ of muons from $Z\rightarrow\mu\mu$ events peak at 1. All
entries exceeding the range of the histogram are added to the last
bin.}

\end{figure}

Because we distinguish CMLLPs solely by their speed$\:\beta$ and
$dE/dx$, we must measure these values for each muon candidate as accurately
as possible. Muons from $Z\rightarrow\mu\mu$ events studied throughout
the data sample allow calibration of the time measurement to better
than 1 ns, with resolutions between 2-4 ns, and to maintain the mean
$dE/dx$ constant to within 2\% over the data-taking period. From
a specific muon scintillation counter we calculate a particle's speed
from the time recorded and the counter's distance from the production
point, and we compute an overall speed from the weighted average of
these individual speeds, using measured resolutions. The ionization
loss data from the typically 8-10 individual energy deposits in the SMT are combined
using an algorithm that omits the largest deposit to reduce the effect
of the Landau tail and corrects for track crossing angle. We normalize
the $dE/dx$ measurements by requiring the $dE/dx$ distribution of muons from $Z\rightarrow\mu\mu$
events to have a maximum at 1. Figure \ref{fig:beta_dedx} shows the distributions
in $\:\beta$ and $dE/dx$ for data and background, which pass the selection criteria described below.

%
{}

The selection of a candidate CMLLP occurs in several steps. Because of the high $p\bar{p}$ collision rate, we employ
a three-level trigger system to reduce the event rate to the 200 Hz
that can be recorded. The trigger system bases its decisions on characteristics
of the event, which for the CMLLP candidates is the presence of a
muon with a high momentum transverse to the beam direction $(p_{T})$.
In order to reduce triggers on cosmic rays, there is a time window
at the initial trigger level. This trigger gate reduces the trigger
efficiency by 10\% for CMLLPs with a mass of 300 GeV (as they will
be slow and some will be out-of-time) and contributes significantly
to our overall acceptance. We avoid a tighter timing gate usually
imposed at the second level of the muon trigger by accepting an alternative
requirement that the muon have a matching track in the SMT.

%
{}

In the standard D0 event reconstruction CMLLPs would appear as muons, which has been verified in detail using MC simulations.
Thus we select events with at least one well
identified high $p_{T}$ muon. For a useful $\:\beta$ measurement,
the event must have scintillator hits in the A-layer and either
the B- or C-layer; and for valid $dE/dx$ data, we require at least
three hits in the SMT. For an optimal tracking and momentum measurement
we require the muon to be central, i.e., with a pseudorapidity$\,$\cite{eta defined}$\:$
$ $$\mid\eta\mid<1.6$$\,$. To reject muons from meson decays we
impose the isolation requirement that the sum of the $p_{T}$ be less
than 2.5 GeV for all other tracks in a cone of radius$\:${\cal R}
$=\sqrt{(\Delta\phi)^{2}+(\Delta\eta)^{2}}$ < 0.5. We also require
that the total transverse calorimeter energy in an annulus of radius
$\:$$0.1<\:${\cal R}$\:<0.4$ about the muon direction be less
than 2.5 GeV. A requirement that the z-coordinate of the muon track 
at the location of closest approach to the beam
axis be $<40$ cm ensures that the particle passes through the SMT.

We impose further criteria to eliminate cosmic rays. To select muons traveling
outwards from the apparent interaction point, we require that its
C-layer time be significantly greater than its A-layer time. We require
also that the muon's distance-of-closest-approach to the beam line
be less than 0.02 cm. These criteria are also applied to a second
muon, if any, in the event. In addition, for events with two muons we require
that the absolute value of the difference between each muon's A-layer
times be less than 10 ns. To reject cosmics that appear as two back-to-back
muons, we require for their pseudo-acolinearity $\Delta\alpha=|\Delta\theta+\Delta\phi-2\pi|>0.05$.

Events with a muon from a $W$ boson decay, with mismeasurements providing
inaccurate values of the muon's $\:\beta$ and $dE/dx$, constitute
a potentially large background. To study selection criteria for CMLLPs,
we calculate the transverse mass M$_{\text{T}}$ \cite{transverse mass}
and select data with M$_{\text{T}}$ < 200 GeV to model the data in
the absence of signal $\:$\cite{MT cut for Ws}. We choose selection
criteria that minimize the number of events surviving from this background
sample compared to events from simulations of the CMLLP signal. We require
that events contain at least one muon with $p_{T}>60$ GeV.  From a separate sample of muons  from $Z\rightarrow\mu\mu$ decays we observe that the association of a spurious scintillator hit can result in an anomalously
low $\:\beta$ value.  We use an algorithm that discards such hits through
minimizing the $\chi^{2}/d.o.f.$ for the $\:\beta$ calculated from
the different scintillator layers. By comparing the effect on the
background sample with the effect on simulated signal, we choose to
eliminate events unless the minimized speed $\chi^{2}/d.o.f.$ < 2. Finally,
we compare the muon's track direction measured by the muon system
with that measured in the central tracker, and eliminate events
with clearly mismatched tracks.

%
{}

To simulate signal events, we generate CMLLP candidates
using {\sc pythia} ~\cite{Pythia}. The long-lived
stop quarks are hadronized using$\,$ \cite{stop-code}.
Because the signature of the CMLLP cascade decays is model dependent
and difficult to simulate accurately, we generate direct pair-production
of the CMLLPs, without including cascade decays. We use the full D0
detector {\sc geant}  \cite{geant}$\:$ simulation to determine the
detector response for these samples (which include overlaid data-based
$p\bar{p}$$ $ interactions). The results are applicable to models
with pair-produced CMLLPs with similar kinematics.

The stop quarks are distinct since they appear in charged or uncharged stop hadrons, which may flip their charge as they pass through the detector. In the
simulation approximately 60\% of stop hadrons are charged following
initial hadronization$\,$\cite{stop-fairbairn}, i.e.,
84\% of the events will have at least one charged stop hadron. Further,
stop hadrons may flip their charge through nuclear interactions as
they pass through material. We assume that stop hadrons have a probability
of $ $2/3 of being charged after multiple nuclear interactions and anti-stop
hadrons, a probability of 1/2 of being charged, consistent with the numbers of possible stop and anti-stop hadronic final states \cite{stop-mackeprangR,stop-mackeprangM,stop-mackeprang}.
For this analysis we require the stop hadron to be charged before and
after passing through the calorimeter, i.e., to be detected both in
the tracker and in the A-layer, and to be charged after the toroid,
i.e., to be detected in the B- or C-layers. The probability for at
least one of the stop hadrons being detected is then 38\%, or 84\%
if charge flipping does not occur. We include these numbers as normalization
factors in the confidence level analysis discussed below.

%
{}

Our final selection criteria is that the candidate's speed $\beta<1.$
Thus, we describe the background by the $\beta<1$ data events with
$M_{T}<200$ GeV, and search for CMLLP candidates in $\beta<1$ data
with $M_{T}>200$ GeV. We normalize the background and data samples
in the $\beta>1$ region, where the contribution of signal is negligible. Because the uncertainties in
the speed measurements depend on the particle's $\eta$, due to detector
geometry, and the distributions in $\eta$ of the muons in the $M_{T}<200$
GeV sample differs from those in the $M_{T}>200$ GeV sample, we use
the signal-free region to derive correction factors for the background sample
that match its $\eta$ distribution to that of the data.

%
{}

We utilize a boosted decision tree (BDT) $\,$\cite{TMVA} to discriminate
signal from background. The most discriminating variables are the CMLLP
candidate's $\beta$ and $dE/dx$, but we also include several related variables:
the speed significance, defined as $(1-\beta)/\sigma_{\beta}$,
the corresponding number of scintillator hits, the energy loss significance
defined as $(dE/dx-1)/\sigma_{dE/dx}$, and the number of SMT hits.
For each mass point in all three signal models we train the BDT with
the signal MC and the background, and then apply it to the data samples.
Figure \ref{fig:beta_dedx} shows the distributions in $\beta$ and
in $dE/dx$ for the data and background samples, as well as for two representative
signals. 

Systematic uncertainties are studied by applying variations to the
background and signal samples and determining the deviations
in the BDT output distributions. Two of the systematic uncertainties affect the shape of
the BDT distribution and their effect is taken into account explicitly in
the limit calculation: the uncertainty due to the width of the Level
1 trigger gate and the uncertainty in the corrections to the MC time resolution; both
are applied only to signal. By examining the signal-like region of
the BDT distributions, we find that the maximum (average) uncertainty
is 10\% (4\%) for the trigger gate width, and 38\% (7\%) for the time resolution
correction.  All other systematic uncertainties affect only the normalization of the BDT output.
The systematic uncertainties on the background are due to the $dE/dx$
modeling (< 0.1\%) and the background normalization, from the specific values used for the $\beta$ (7.2\%) and $M_{T}$ requirements (2.2\%). The systematic
uncertainties on the signal include muon identification (2\%) and
the integrated luminosity (6.1\%) \cite{d0lumi}. The systematic uncertainties
associated with the corrections to the muon $p_{T}$$ $ resolution and to the $dE/dx$ resolution, as well as  the choice of PDF and factorization scale, are all below 1\%.

%

%
\begin{figure}[]

\noindent \begin{flushleft}
\includegraphics[scale=0.47]{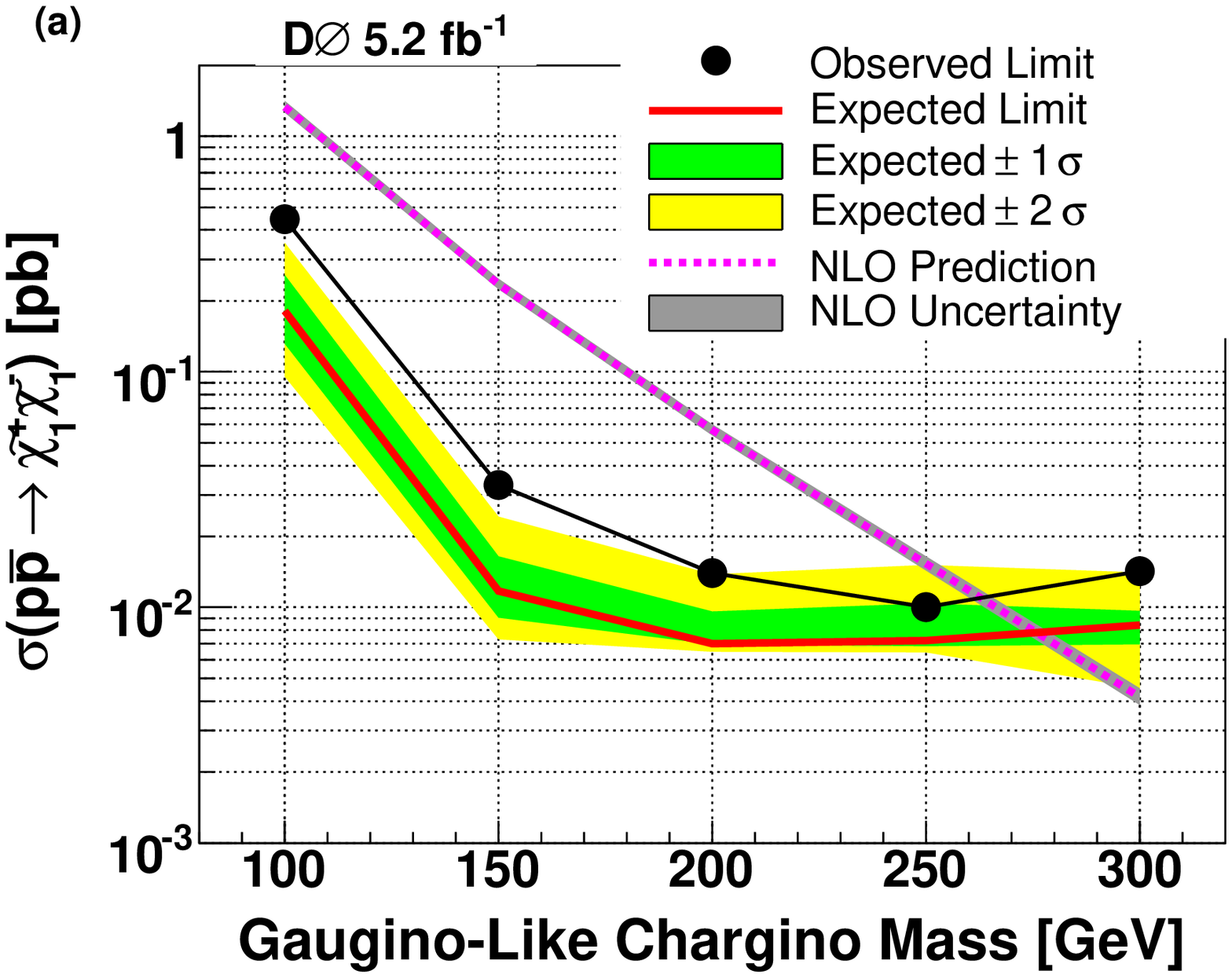}
\par\end{flushleft}

\vspace{-0.4 cm}

\noindent \begin{flushleft}
\includegraphics[scale=0.47]{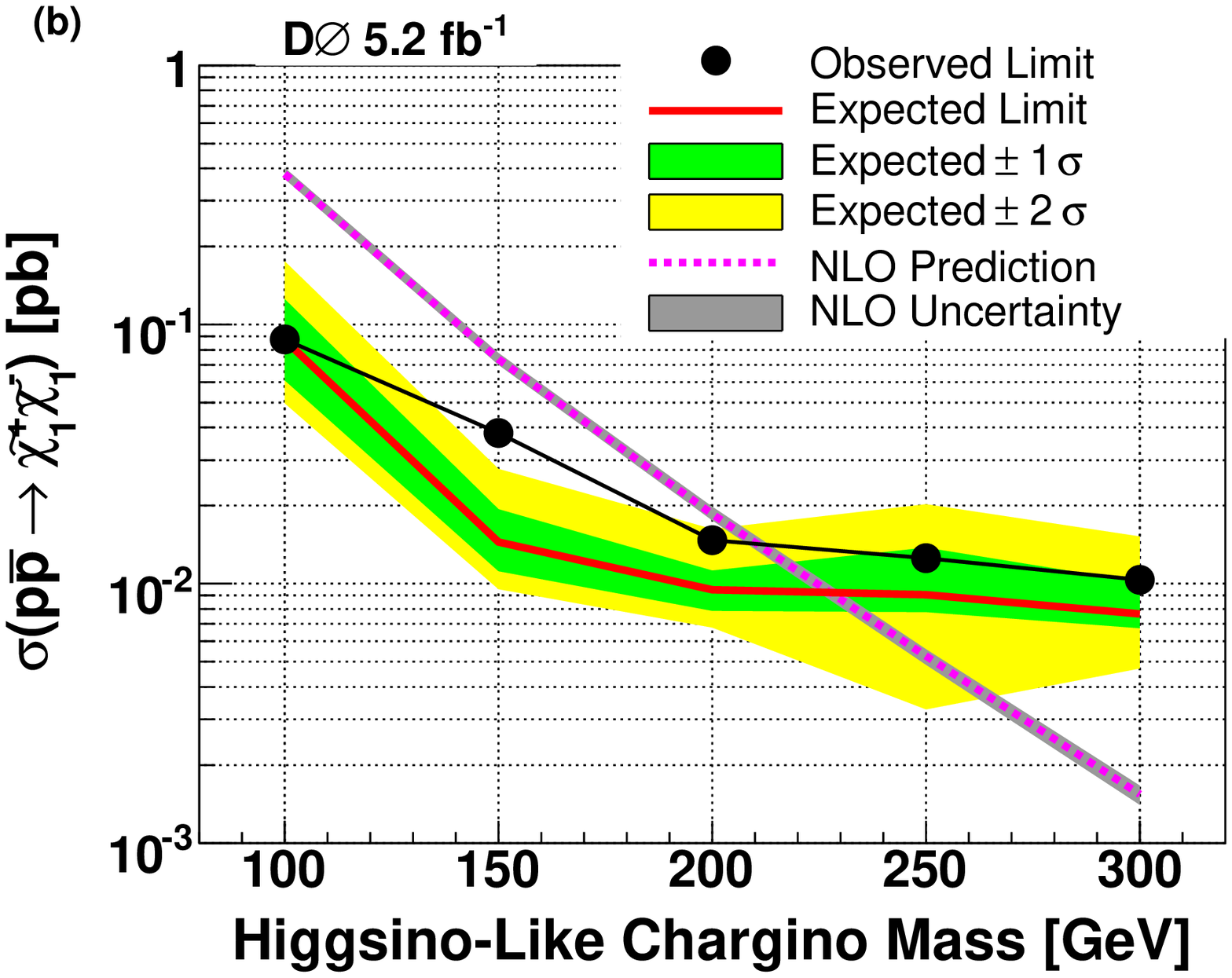}
\par\end{flushleft}

\vspace{-0.4 cm}

\noindent \begin{flushleft}
\includegraphics[scale=0.47]{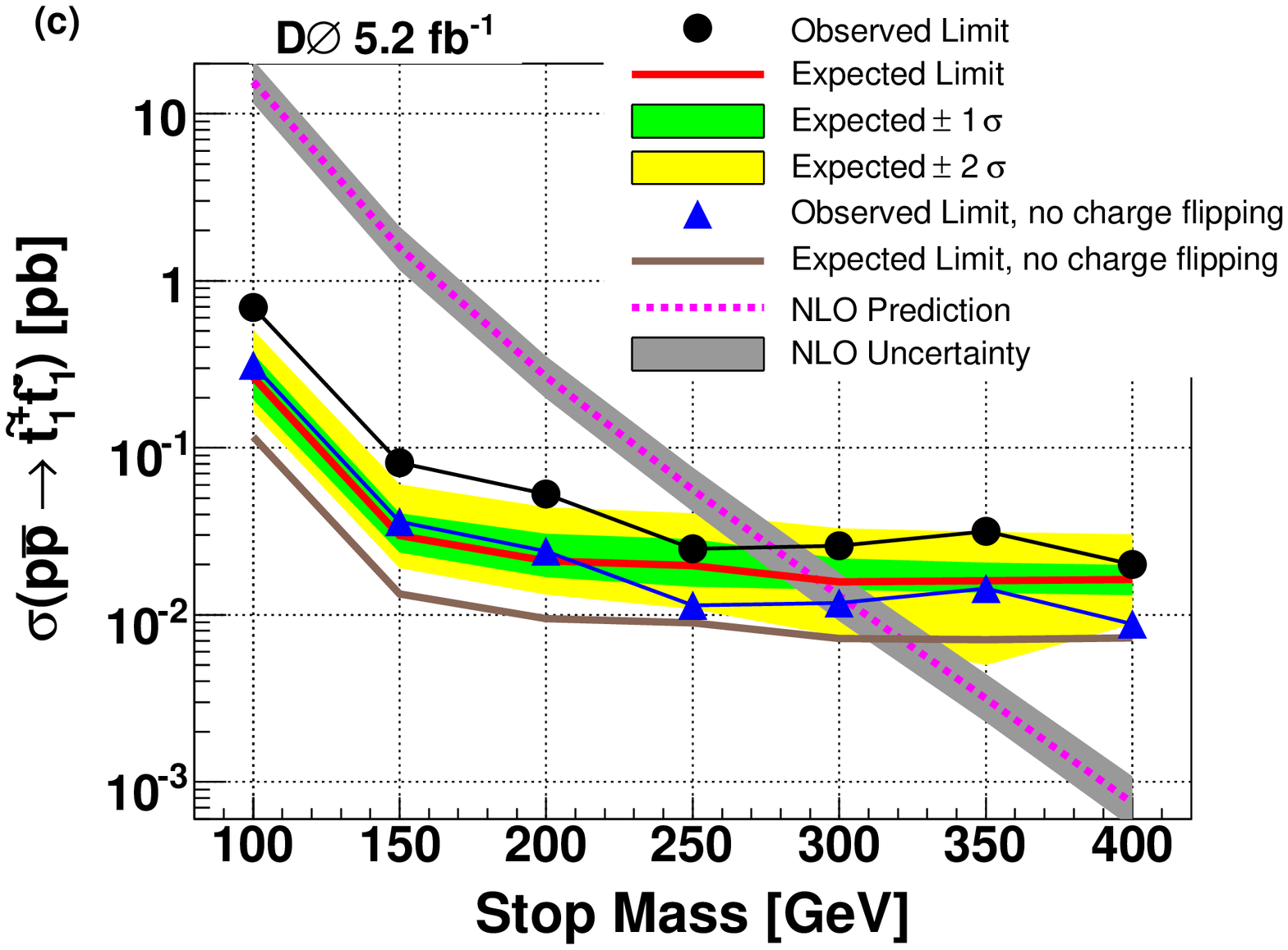}
\par\end{flushleft}

\vspace{-0.6 cm}

\caption{\label{fig:limits}95\% C.L. cross-section limits as a function of
mass for (a) gaugino-like charginos, (b) higgsino-like charginos, and
(c) stop quarks. The stop quark limits are displayed
for the assumed charge flipping (charge survival probability = 38\%)
and for no charge flipping (charge survival probability = 84\%).}

\end{figure}

We obtain the 95\% C.L. cross section limits from the BDT output distributions, constraining systematic uncertainties to data in background dominated 
regions $\,$\cite{Collie}.
These limits are shown in Figure \ref{fig:limits}, together with
the NLO theoretical signal cross sections, computed with {\sc prospino}
~\cite{Prospino}. The limits are of order 0.01 pb for directly
pair-produced CMLLPs with masses between 200 and 300 GeV.  
Using the theoretical cross sections,
we are able to exclude gaugino-like charginos below 267 GeV
and higgsino-like charginos below 217 GeV. For stop quarks,
we assume a charge survival probability of 38\%, as discussed above,
and exclude masses below 285 GeV. If charge flipping does not
occur, we would obtain a higher mass limit, as indicated in Fig. 
\ref{fig:limits}c.

As seen in Figure \ref{fig:limits}, the observed
limit exceeds the expected limit at various mass points by as much as 2.5
standard
deviations, for all signals tested, due to the presence of the same few data
events with high BDT discriminant values.  This discrepancy reflects the
excesses of data compared to background observed in Fig. \ref{fig:beta_dedx} for the
distributions both in beta (around
0.6) and dE/dx (around 2.8).  We have compared these events with background
events and conclude that the
observed excess is consistent with a statistical fluctuation in the number
of such events.
%
{}

In summary, we have performed a search for charged, massive long-lived
particles using $5.2$ fb$^{\text{-1}}$ of integrated luminosity
with the D0 detector. We find no evidence of signal and set 95\% C.L.
cross-section upper limits of order 0.01 pb for pair-produced CMLLPs of mass 200-300 GeV. At 95\% C.L. we exclude pair-produced long-lived stop quarks
with mass below 285 GeV, gaugino-like charginos below 267 GeV, and
higgsino-like charginos below 217 GeV. These are presently the most
restrictive limits for chargino CMLLPs, with about a factor of five
improvement over the previous D0 cross section limits $\,$\cite{D0 MSP PRL}.

%
{}

We thank the staffs at Fermilab and collaborating institutions, and
acknowledge support from the DOE and NSF (USA); CEA and CNRS/IN2P3
(France); FASI, Rosatom and RFBR (Russia); CNPq, FAPERJ, FAPESP and
FUNDUNESP (Brazil); DAE and DST (India); Colciencias (Colombia); CONACyT
(Mexico); KRF and KOSEF (Korea); CONICET and UBACyT (Argentina); FOM
(The Netherlands); STFC and the Royal Society (United Kingdom); MSMT
and GACR (Czech Republic); CRC Program and NSERC (Canada); BMBF and
DFG (Germany); SFI (Ireland); The Swedish Research Council (Sweden);
and CAS and CNSF (China).

\end{document}

%% file: author_list.tex
\affiliation{Universidad de Buenos Aires, Buenos Aires, Argentina}
\affiliation{LAFEX, Centro Brasileiro de Pesquisas F{\'\i}sicas, Rio de Janeiro, Brazil}
\affiliation{Universidade do Estado do Rio de Janeiro, Rio de Janeiro, Brazil}
\affiliation{Universidade Federal do ABC, Santo Andr\'e, Brazil}
\affiliation{Instituto de F\'{\i}sica Te\'orica, Universidade Estadual Paulista, S\~ao Paulo, Brazil}
\affiliation{University of Science and Technology of China, Hefei, People's Republic of China}
\affiliation{Universidad de los Andes, Bogot\'{a}, Colombia}
\affiliation{Charles University, Faculty of Mathematics and Physics, Center for Particle Physics, Prague, Czech Republic}
\affiliation{Czech Technical University in Prague, Prague, Czech Republic}
\affiliation{Center for Particle Physics, Institute of Physics, Academy of Sciences of the Czech Republic, Prague, Czech Republic}
\affiliation{Universidad San Francisco de Quito, Quito, Ecuador}
\affiliation{LPC, Universit\'e Blaise Pascal, CNRS/IN2P3, Clermont, France}
\affiliation{LPSC, Universit\'e Joseph Fourier Grenoble 1, CNRS/IN2P3, Institut National Polytechnique de Grenoble, Grenoble, France}
\affiliation{CPPM, Aix-Marseille Universit\'e, CNRS/IN2P3, Marseille, France}
\affiliation{LAL, Universit\'e Paris-Sud, CNRS/IN2P3, Orsay, France}
\affiliation{LPNHE, Universit\'es Paris VI and VII, CNRS/IN2P3, Paris, France}
\affiliation{CEA, Irfu, SPP, Saclay, France}
\affiliation{IPHC, Universit\'e de Strasbourg, CNRS/IN2P3, Strasbourg, France}
\affiliation{IPNL, Universit\'e Lyon 1, CNRS/IN2P3, Villeurbanne, France and Universit\'e de Lyon, Lyon, France}
\affiliation{III. Physikalisches Institut A, RWTH Aachen University, Aachen, Germany}
\affiliation{Physikalisches Institut, Universit{\"a}t Freiburg, Freiburg, Germany}
\affiliation{II. Physikalisches Institut, Georg-August-Universit{\"a}t G\"ottingen, G\"ottingen, Germany}
\affiliation{Institut f{\"u}r Physik, Universit{\"a}t Mainz, Mainz, Germany}
\affiliation{Ludwig-Maximilians-Universit{\"a}t M{\"u}nchen, M{\"u}nchen, Germany}
\affiliation{Fachbereich Physik, Bergische Universit{\"a}t Wuppertal, Wuppertal, Germany}
\affiliation{Panjab University, Chandigarh, India}
\affiliation{Delhi University, Delhi, India}
\affiliation{Tata Institute of Fundamental Research, Mumbai, India}
\affiliation{University College Dublin, Dublin, Ireland}
\affiliation{Korea Detector Laboratory, Korea University, Seoul, Korea}
\affiliation{CINVESTAV, Mexico City, Mexico}
\affiliation{Nikhef, Science Park, Amsterdam, the Netherlands}
\affiliation{Radboud University Nijmegen, Nijmegen, the Netherlands and Nikhef, Science Park, Amsterdam, the Netherlands}
\affiliation{Joint Institute for Nuclear Research, Dubna, Russia}
\affiliation{Institute for Theoretical and Experimental Physics, Moscow, Russia}
\affiliation{Moscow State University, Moscow, Russia}
\affiliation{Institute for High Energy Physics, Protvino, Russia}
\affiliation{Petersburg Nuclear Physics Institute, St. Petersburg, Russia}
\affiliation{Instituci\'{o} Catalana de Recerca i Estudis Avan\c{c}ats (ICREA) and Institut de F\'{i}sica d'Altes Energies (IFAE), Barcelona, Spain}
\affiliation{Stockholm University, Stockholm and Uppsala University, Uppsala, Sweden}
\affiliation{Lancaster University, Lancaster LA1 4YB, United Kingdom}
\affiliation{Imperial College London, London SW7 2AZ, United Kingdom}
\affiliation{The University of Manchester, Manchester M13 9PL, United Kingdom}
\affiliation{University of Arizona, Tucson, Arizona 85721, USA}
\affiliation{University of California Riverside, Riverside, California 92521, USA}
\affiliation{Florida State University, Tallahassee, Florida 32306, USA}
\affiliation{Fermi National Accelerator Laboratory, Batavia, Illinois 60510, USA}
\affiliation{University of Illinois at Chicago, Chicago, Illinois 60607, USA}
\affiliation{Northern Illinois University, DeKalb, Illinois 60115, USA}
\affiliation{Northwestern University, Evanston, Illinois 60208, USA}
\affiliation{Indiana University, Bloomington, Indiana 47405, USA}
\affiliation{Purdue University Calumet, Hammond, Indiana 46323, USA}
\affiliation{University of Notre Dame, Notre Dame, Indiana 46556, USA}
\affiliation{Iowa State University, Ames, Iowa 50011, USA}
\affiliation{University of Kansas, Lawrence, Kansas 66045, USA}
\affiliation{Kansas State University, Manhattan, Kansas 66506, USA}
\affiliation{Louisiana Tech University, Ruston, Louisiana 71272, USA}
\affiliation{Boston University, Boston, Massachusetts 02215, USA}
\affiliation{Northeastern University, Boston, Massachusetts 02115, USA}
\affiliation{University of Michigan, Ann Arbor, Michigan 48109, USA}
\affiliation{Michigan State University, East Lansing, Michigan 48824, USA}
\affiliation{University of Mississippi, University, Mississippi 38677, USA}
\affiliation{University of Nebraska, Lincoln, Nebraska 68588, USA}
\affiliation{Rutgers University, Piscataway, New Jersey 08855, USA}
\affiliation{Princeton University, Princeton, New Jersey 08544, USA}
\affiliation{State University of New York, Buffalo, New York 14260, USA}
\affiliation{Columbia University, New York, New York 10027, USA}
\affiliation{University of Rochester, Rochester, New York 14627, USA}
\affiliation{State University of New York, Stony Brook, New York 11794, USA}
\affiliation{Brookhaven National Laboratory, Upton, New York 11973, USA}
\affiliation{Langston University, Langston, Oklahoma 73050, USA}
\affiliation{University of Oklahoma, Norman, Oklahoma 73019, USA}
\affiliation{Oklahoma State University, Stillwater, Oklahoma 74078, USA}
\affiliation{Brown University, Providence, Rhode Island 02912, USA}
\affiliation{University of Texas, Arlington, Texas 76019, USA}
\affiliation{Southern Methodist University, Dallas, Texas 75275, USA}
\affiliation{Rice University, Houston, Texas 77005, USA}
\affiliation{University of Virginia, Charlottesville, Virginia 22901, USA}
\affiliation{University of Washington, Seattle, Washington 98195, USA}
\author{V.M.~Abazov} \affiliation{Joint Institute for Nuclear Research, Dubna, Russia}
\author{B.~Abbott} \affiliation{University of Oklahoma, Norman, Oklahoma 73019, USA}
\author{B.S.~Acharya} \affiliation{Tata Institute of Fundamental Research, Mumbai, India}
\author{M.~Adams} \affiliation{University of Illinois at Chicago, Chicago, Illinois 60607, USA}
\author{T.~Adams} \affiliation{Florida State University, Tallahassee, Florida 32306, USA}
\author{G.D.~Alexeev} \affiliation{Joint Institute for Nuclear Research, Dubna, Russia}
\author{J.~Alimena} \affiliation{Brown University, Providence, Rhode Island 02912, USA}
\author{G.~Alkhazov} \affiliation{Petersburg Nuclear Physics Institute, St. Petersburg, Russia}
\author{A.~Alton$^{a}$} \affiliation{University of Michigan, Ann Arbor, Michigan 48109, USA}
\author{G.~Alverson} \affiliation{Northeastern University, Boston, Massachusetts 02115, USA}
% ...
%
%
\collaboration{The D0 Collaboration\footnote{with visitors from
$^{a}$Augustana College, Sioux Falls, SD, USA,
$^{b}$The University of Liverpool, Liverpool, UK,
$^{c}$UPIITA-IPN, Mexico City, Mexico,
$^{d}$SLAC, Menlo Park, CA, USA,
$^{e}$University College London, London, UK,
$^{f}$Centro de Investigacion en Computacion - IPN, Mexico City, Mexico,
$^{g}$ECFM, Universidad Autonoma de Sinaloa, Culiac\'an, Mexico,
and 
$^{h}$Universit{\"a}t Bern, Bern, Switzerland.
$^{\ddag}$Deceased.
}} \noaffiliation
\vskip 0.25cm

%% file: prl_CMLLP_V81.bbl
\begin{thebibliography}{30}
\bibitem{eta defined} The D0 coordinate system is cylindrical with
the $z$-axis along the proton beam direction, and the polar and azimuthal
angles are denoted by $\theta$ and $\phi$, respectively. The pseudorapidity
is defined as $\eta=$-ln{[}tan$\left(\theta/2\right)${]}.
\bibitem{transverse mass} The transverse mass is defined by $M_{T}=\sqrt{(E_{T}+{\not}E_{T})^{2}-(p_{x}+{\not}E_{x})^{2}-(p_{y}+{\not}E_{y})^{2}}$,
where $E_{T}$ is the total energy transverse to the axis of the colliding
beams, and ${\not}E_{T}$ is the total unbalanced or missing transverse
energy.  The $E_{T}$ and ${\not}E_{T}$ are measured using calorimeter deposits corrected for identified jets and leptons.
\bibitem{MT cut for Ws} The requirement $M_{T}<200$ GeV is customarily
used to select $W$ events.  There is negligible contamination of potential signal events in this data sample.
\end{thebibliography}
